\documentclass{iopart}
\usepackage[utf8]{inputenc}
\usepackage{bm}
\usepackage[singlespacing]{setspace} 
\usepackage{indentfirst}
\usepackage{graphicx}
\usepackage{color}

\newcommand{\Alfven}{Alfv\'{e}n }

\newcommand{\pper}{{p_\perp}}
\newcommand{\ppar}{{p_\parallel}}
\newcommand{\ch}{{\hat{c}}}

\newcommand{\ten}[1]{\overline{\bm{#1}}}

\newcommand{\bPsi}{\Psi_N}
\newcommand{\gparone}{\gamma_{\parallel 1}}
\newcommand{\gpartwo}{\gamma_{\parallel 2}}
\newcommand{\gperone}{\gamma_{\perp 1}}
\newcommand{\gpertwo}{\gamma_{\perp 2}}

\newcommand{\eqref}[1]{Eq. (\ref{#1})}
\newcommand{\Eqref}[1]{\Eref{#1}}
\newcommand{\figref}[1]{Fig.\ref{#1}}
\newcommand{\Figref}[1]{\Fref{#1}}

\newcommand{\secref}[1]{Section \ref{#1}}
\newcommand{\rcite}[1]{Ref. \cite{#1}}

\begin{document}
\title{Modeling the effect of anisotropic pressure on tokamak plasmas normal modes and continuum using fluid approaches}
\author{Z.S. Qu$^1$, M.J. Hole$^1$ and M. Fitzgerald$^{2}$}
\address{$^1$ Research School of Physics and Engineering, the Australian National University, Canberra ACT 0200, Australia}
\address{$^2$ EURATOM/CCFE Fusion Association, Culham Science Centre, Abingdon, Oxon, OX14 3DB, UK.}
\ead{\mailto{zhisong.qu@anu.edu.au}}

\begin{abstract}
Extending the ideal MHD stability code MISHKA, 
a new code, MISHKA-A, is developed to study the impact of pressure anisotropy on plasma stability.
Based on full anisotropic equilibrium and geometry,
the code can provide normal mode analysis with three fluid closure models:
the single adiabatic model (SA), the double adiabatic model (CGL) and the incompressible model.
A study on the plasma continuous spectrum shows that in low beta, large aspect ratio plasma,
the main impact of anisotropy lies in the modification of the BAE gap and the sound frequency, if the $q$ profile is conserved.
The SA model preserves the BAE gap structure as ideal MHD,
while in CGL the lowest frequency branch does not touch zero frequency at the resonant flux surface where $m+nq=0$, 
inducing a gap at very low frequency.
Also, the BAE gap frequency with bi-Maxwellian distribution in both model becomes higher if $\pper > \ppar$ with a $q$ profile dependency.
As a benchmark of the code, we study the $m/n=1/1$ internal kink mode.
Numerical calculation of the marginal stability boundary with bi-Maxwellian distribution shows a good agreement with the generalized incompressible Bussac criterion
[A. B. Mikhailovskii, Sov. J. Plasma Phys 9, 190 (1983)]:
the mode is stabilized(destabilized) if $\ppar < \pper$ ($\ppar > \pper$).
\end{abstract}

\section{Introduction}
The magnetohydrodynamics (MHD) theory is widely applied in fusion plasma,
providing a great aid in explaining various plasma instabilities and the plasma oscillating spectra below the ion cyclotron frequency.
In modern toroidal magnetic confinement devices, the plasma contains significant fast populations originated from neutral beam injection (NBI) and ion cyclotron resonance heating (ICRH),
inducing strong pressure anisotropy \cite{Fasoli2007}.
The magnitude of anisotropy can reach $\ppar \approx 1.7 \pper$ in a MAST beam heated discharge \cite{Hole2011,  Qu2014},
or $\pper \approx 2.5 \ppar$ in a JET ICRH discharge \cite{Zwingmann2001},
with $\ppar$ and $\pper$ the pressure parallel and perpendicular to the magnetic field lines, respectively.
However, the physics of pressure anisotropy is not covered by the isotropic MHD theory.

In the regime where wave-particle interaction is not important,
a fluid approach is often used with a reasonable fluid closure (like the adiabatic condition for ideal MHD)
for phenomena only related to the macroscopic quantities such as density, current and pressure.
Many attempts have be made to incorporate anisotropy into the fluid theory.
Chew, Goldberger and Low (CGL) \cite{Chew1956} first introduced the widely-applied form of pressure tensor and derived the double-adiabatic (CGL) closure, 
with its energy principle later derived by Bernstein \etal \cite{Bernstein1958}.
Unlike MHD, CGL assumes parallel and perpendicular pressures doing work independently in a collisionless plasma, 
therefore cannot reduce to MHD in the isotropic limit.
It was found that CGL overestimates $\delta W$, the perturbed potential energy, compared to the kinetic theory,
while MHD underestimates it \cite{Kruskal1958,Rosenbluth1959}.
Also, the mirror stability limit given by CGL does not match the result of kinetics theory \cite{Kato1966, Tajiri1967}.
The major problem with CGL comes from the ignored heat flow when the mode frequency is comparable or smaller than the particle streaming frequency, 
especially in the vicinity of marginal stability boundary \cite{Kulsrud1983, Chust2006}.
Still, the CGL closure is implemented in many stability treatments, such as the ballooning modes \cite{Cooper1981, Wang1990}.
To overcome these drawbacks of CGL, some authors have proposed alternative fluid closures,
for instance the double polytropic laws \cite{Hau1993}, a higher-order-momentum closure \cite{Ramos2003,Ramos2005},
and recently, the single adiabatic (SA) model \cite{Fitzgerald2015} which has the unique property of producing the same results as the MHD model for isotropic equilibria.
Another pathway to overcome the drawbacks of CGL is to use hybrid approaches, 
in which thermal components are described by MHD and the fast ions by kinetics.
The impact of pressure anisotropy is often investigated using kinetics energy princples \cite{Kruskal1958,Rosenbluth1959,Antonsen1982}.
In tokamaks, efforts have been made to study sawtooth modes (see Graves \etal \cite{Graves2005} and Chapman \etal \cite{Chapman2011} and references therein)
and interchange modes \cite{Connor1976}.
There are also significant developments in stellarators.
The ANIMEC code \cite{Cooper2009} solves the 3D anisotropic equilibrium with the fast ion described by a guiding center distribution function,
and is further applied to model anisotropy on LHD\cite{Asahi2013}.
An energy principle which assumes non-interacting hot particles \cite{Johnson1969}
is implemented in the ideal MHD code TERPSICHORE \cite{Anderson1990}
to model anisotropic-pressure interchange modes in a beam heated LHD discharge \cite{Cooper2012}.
Despite its shortcoming, the fluid approach can aid in the understanding of various effects due to its simple and intuitive nature.
To date, there are few numerical studies on the oscillating spectrum of a toroidal anisotropic plasma.

In the regime where significant wave-particle resonance exists, a pertubtive approach, 
in which the equilibrium and the linear mode eigenfunctions are modeled by fluid theory and the wave-particle interaction by kinetic theory, 
is widely implemented.
In tokamaks,
one of the most utilized tool chains is the HELENA-MISHKA-HAGIS combination \cite{Huysmans1991,Mikhailovskii1997,Pinches1998},
with the equilibrium, geometry and mode eigenfunctions calculated by ideal MHD,
while the fast ion response and non-linear mode evolution are described by drift-kinetics equations.
It has been successful in resolving the fast-particle-excited global \Alfven eigenmodes
(see reviews \cite{Sharapov2013, Hole2014} and references therein).
Recently, several equilibrium codes \cite{Qu2014, Guazzotto2004, Fitzgeral2013} have been developed to study the equilibrium of anisotropic and toroidally rotating plasmas.
For linear stability problem, efforts have been made to include the physics of diamagnetic drift and toroidal flow 
into MISHKA \cite{Huysmans2001, Chapman2006} for an isotropic equilibrium,
while the impact of pressure anisotropy based on a full anisotropic equilibrium and geometry remains untouched.
Our previous study using current remapping techniques shows that anisotropy can modify the $q$ profile in MAST,
inducing double TAE modes with different localization \cite{Hole2011, Hole2013}, and thus a double wave-particle resonance.
This also serves as a motivation to develop a MISHKA-like code to study the impact of anisotropy on linear stability,
meanwhile drive a kinetic code using a fully anisotropic framework.

This work is organized as follows. 
In \secref{sec:plasma_model}, we state our basic assumptions and list the plasma equations used in the paper.
\secref{sec:equilibrium_geometry} briefly describes the anisotropic equilibria and introduce the straight field line coordinates,
serving as a base point for the stability treatment.
Then in \secref{sec:perturbed_eqs}, we derive the linearized momentum equation, ideal Ohmic's law and the fluid closure equations
which are ready to use in a MISHKA-like numerical code.
\secref{sec:numerical} introduces the implementation of the derived equations into a global normal mode code, MISHKA-A,
and a continuous spectrum code, CSMISH-A.
Using these tools, we study the impact of anisotropy on the plasma continuous spectrum and the internal kink mode,
shown in \secref{sec:continuum} and \secref{sec:internal_kink}, respectively.
We also compare the numerical results with existing analytical theory,
serving as a code benchmark.
Finally, \secref{sec:conclusion} summarizes the paper and draws the conclusion.

\section{Plasma Model} \label{sec:plasma_model}
We start from a plasma described by the first two moments of the Vlasov Equation (the continuity and the momentum equation),
the Maxwell Equations and the ideal Ohmic law.
The basic equations are
\begin{equation}
 \frac{d \rho}{dt} + \rho (\nabla \cdot \bm{V}) = 0,
 \label{eq:continuity}
\end{equation}
\begin{equation}
 \rho \frac{\partial\bm{V}}{\partial t} = -\nabla \cdot \ten{P} + \bm{j} \times \bm{B},
 \label{eq:gcp_momentum}
\end{equation}
\begin{equation}
 \frac{\partial \bm{B}}{\partial t} = \nabla \times (\bm{V} \times \bm{B}),
 \label{eq:maxwell_1}
\end{equation}
\begin{equation}
\bm{j} = \nabla \times \bm{B}, 
\label{eq:maxwell_2}
\end{equation}
\begin{equation}
 \nabla \cdot \bm{B} = 0, 
 \label{eq:maxwell_3}
\end{equation}
where $\rho$ is the mass density, $\bm{V}$ the mass velocity, $\ten{P}$ the second rank pressure tensor,
$\bm{j}$ the current density and $\bm{B}$ the magnetic field.
For simplicity, we use a natural MHD unit system where $\mu_0$, the vacuum permeability, is set to $1$.
All electromagnetic fields, fluxes and vector potentials can be restored to S.I. units
with a transformation $\dots \rightarrow \dots/ \sqrt{\mu_0}$ (e.g. $B \rightarrow B/ \sqrt{\mu_0}$)
and all currents with $j \rightarrow \sqrt{\mu_0} j$.
\Eqref{eq:continuity} is the continuity equation.
\Eqref{eq:gcp_momentum} is the momentum equation.
\Eqref{eq:maxwell_1}, (\ref{eq:maxwell_2}) and (\ref{eq:maxwell_3}) are the Maxwell Equations with ideal Ohmic law ignoring the displacement currents.
The pressure tensor $\ten{P}$ takes the CGL form, i.e.
\begin{equation}
 \ten{P} = \pper \ten{I} + \Delta \bm{B}\bm{B}, \quad
 \Delta =  \frac {\ppar - \pper} {B^2},
 \label{eq:gcp_pressure}
\end{equation}
with $\ten{I}$ the identity tensor, 
$\pper$ and $\ppar$ the pressure perpendicular and parallel to the magnetic field, respectively.
In our treatment, the finite Larmor radius (FLR) and the finite orbit width (FOW) effects are ignored.
These effects can be important for fast particles, but resolving them requires
FLR correction of non-diagonal pressure tensor terms (such as Chhajlani \etal \cite{Chhajlani1980} for CGL) or kinetics/gyro-kinetics approaches,
which are not considered in this paper.

In this paper, we implement the standard linearization method, 
which expands all quantities into a combination of a time-averaging equilibrium part and a small time-dependent part,
which varies with $e^{\lambda t}$.
The mode frequency $\omega$ and growth rate $\gamma$ are related to $\lambda$ through the relationship $\lambda$ = $\gamma - i \omega$.
By substituting these representatives into the plasma equations and considering the zeroth and the first order separately,
the equations are then converted into a time-independent equilibrium problem and a linearized stability problem.
We drop the subscripts ``0'' for equilibrium quantities for convenience.

To close the set of equations, 
one needs to introduce a ``fluid closure'' which relates $\ppar$ and $\pper$ to other known variables.
In this work, we examine three fluid closures: the single adiabatic model \cite{Fitzgerald2015}, the double adiabatic model \cite{Chew1956},
and the incompressible limit given by Mikhailovskii \cite{Mikhailovskii1983}.
The single adiabatic model serves as a generalization of MHD.
While keeping the adiabaticity assumption of MHD, 
it assumes that the parallel and perpendicular pressure are doing joint work,
and therefore resolves the isotropic part of the pressure perturbation.
This fluid closure equation is given by
\begin{eqnarray}
\fl \frac{\partial \tilde{\ppar}}{\partial t} = \frac{\partial \tilde{\pper}}{\partial t} =
 - \tilde{\bm{V}} \cdot \nabla \left(\frac{1}{3} \ppar + \frac{2}{3} \pper \right) \nonumber\\
 - \left(\frac{1}{3} \ppar + \frac{4}{3} \pper \right) \nabla \cdot \tilde{\bm{V}}
 - \left(\frac{2}{3} \ppar - \frac{2}{3} \pper \right) \bm{b} \cdot (\bm{b} \cdot \nabla \tilde{\bm{V}}),
 \label{eq:pressure_SA}
\end{eqnarray}
in which the unit vector $\bm{b} = \bm{B}/B$ is the direction of the magnetic field line.
In contrast, the double adiabatic model assumes that parallel and perpendicular pressure do adiabatic work independently.
The fluid closure equations, $d/dt(\pper/\rho B) = d/dt(\ppar B^2 / \rho^3) = 0$,
after substituting \eqref{eq:continuity} for $d\rho/dt$ 
and $\bm{B}$ direction of \eqref{eq:maxwell_1} for $dB/dt$, are rewritten as
\begin{eqnarray}
 \frac{\partial \tilde{\ppar}}{\partial t} = - \tilde{\bm{V}} \cdot \nabla \ppar
 - \ppar (\nabla \cdot \tilde{\bm{V}}) - 2 \ppar \bm{b} \cdot (\bm{b} \cdot \nabla \tilde{\bm{V}}),
 \label{eq:pressure_DA1} \\
 \frac{\partial \tilde{\pper}}{\partial t} = - \tilde{\bm{V}} \cdot \nabla \pper
 - 2 \pper (\nabla \cdot \tilde{\bm{V}}) + \pper \bm{b} \cdot (\bm{b} \cdot \nabla \tilde{\bm{V}}).
 \label{eq:pressure_DA2}
\end{eqnarray}
Finally, the incompressible closure is obtained when the Lagrangian perturbed distribution function is set to zero,
i.e. $d\tilde{f}/dt = \partial \tilde{f}/\partial t + \tilde{\bm{V}}_\perp \cdot \nabla f_0 = 0$,
where $\tilde{f}$ is the Euler perturbed distribution function and $f_0$ is the equilibrium distribution function. 
After integrating over the velocity space,
the incompressible closure fluid closure is given by
\begin{eqnarray}
\frac{ \partial \tilde{\ppar}}{\partial t} = - \tilde{V}^1 \left(\frac{\partial{\ppar}}{\partial s}\right)_B,\label{eq:pressure_IC1} \\
\frac{ \partial \tilde{\pper}}{\partial t} = - \tilde{V}^1 \left(\frac{\partial{\pper}}{\partial s}\right)_B,\label{eq:pressure_IC2}
\end{eqnarray}
where $\tilde{V}^1$ is the contravariant component of the straight field line coordinates $(s, \vartheta, \varphi)$,
which will be introduced in the next section.

\section{Equilibrium and geometry}\label{sec:equilibrium_geometry}
For the zeroth order equilibrium problem, the time derivatives $\partial/\partial t = 0$.
In this work, we ignore all equilibrium flows, i.e. $\bm{V}_0 = 0$.
Using \eqref{eq:maxwell_3} in an axisymmetric tokamak geometry,
the equilibrium magnetic field in cylindrical coordinate $(R,Z,\varphi)$ is written as
\begin{equation}
 \bm{B} = \nabla \Psi \times \nabla \varphi + F \nabla \varphi,
 \label{eq:b-definition}
\end{equation}
where $\Psi$ is the poloidal magnetic flux, $F \equiv R B_\varphi$,
and $B_\varphi$ is the toroidal magnetic field.
We note that unlike plasma with isotropic pressure,
we do not require $F$ to be a flux function.

Substituting \eqref{eq:maxwell_2}, (\ref{eq:gcp_pressure}) and (\ref{eq:b-definition}) into \eqref{eq:gcp_momentum},
the component in $\nabla \varphi$ direction gives rise to a flux function $F_M(\Psi) \equiv RB_\varphi(1-\Delta)$,
while the $\nabla \Psi$ direction gives the anisotropy modified Grad-Shafranov Equation (GSE).
The modified GSE have two equivalent forms, 
the pressure form and the enthalpy form (See \cite {Qu2014} and references therein).
In the pressure form of the GSE, the input profiles are specified by $F_M(\Psi)$ and a 2D profile $p_\parallel(\Psi,B)$.
This 2D pressure profile is usually obtained by taking the moments of guiding center distribution functions \cite{Iacono1990}
either analytically \cite{Cooper1980} or numerically \cite{Salberta1987} (see \rcite{Takeda1991} for a brief review).
The enthalpy form of the GSE, when solved assuming the distribution functions are bi-Maxwellian 
and the parallel temperature is a flux function $T_{\parallel}=T_{\parallel}(\Psi)$,
requires four flux functions $\{H,T_{\parallel}, F_M, \Theta\}$ as input,
corresponding to the density, parallel temperature, toroidal field and anisotropy, respectively.
The profile $H (\Psi)$ gives the radial shape of the density profile,
and in isotropic plasma we have $\rho = \exp(H/T_\parallel)$.
The profile $\Theta (\Psi)$ defines the anisotropy magnitude $\pper/\ppar$,
which is given by
\begin{equation}
 \frac{\pper}{\ppar} = \frac{B}{|B-\Theta T_\parallel|}.
\end{equation}
The density and pressures are then linked to these profiles through
\begin{eqnarray}
 \rho = \frac{\pper}{\ppar} \exp{\frac{H}{T_\parallel}}, \quad
 \label{eq:rho_tper}
\end{eqnarray}
and
\begin{equation}
  \ppar = \rho T_\parallel, \quad
 \pper = \rho T_\perp = \rho T_\parallel \frac{B}{|B-\Theta T_\parallel|}.
 \label{eq:ppar_pper}
\end{equation}
These equation are identical to taking the moments of a bi-Maxwellian distribution function of the form in McClements \etal\cite{McClements1996}, written as
\begin{equation}
 F(\mu, E, \Psi) =  n_{r}(\Psi) \frac{A(r)}{\sqrt{2 \pi T_\perp(\Psi)}^{3}} \exp \left[- \frac{|E-\mu B_0|}{T_\parallel(\Psi)} - \frac{\mu B_0}{T_\perp(\Psi)} \right],
\end{equation}
where $A(r)$ is a normalization factor and $\Theta$ is just a convenient representation of the combination
\begin{equation}
 \Theta = \left(\frac{1}{T_\parallel(\Psi)}-\frac{1}{T_\perp(\Psi)} \right) B_0.
\end{equation}
      In this paper, we will use this bi-Maxwellian model to explore the impact of anisotropy on stability,
  since it is the simplest model that captures pressure anisotropy for both ICRH and NBI.
  The model has limitations,
  such that it takes all species as a single bi-Maxwellian therefore cannot reproduce the long tail of ICRH fast ions,
  and that it omits any physics due to fine structure of pitch angle dependency of the distribution function 
  (i.e. non-bi-Maxwellian structure).
 However, it does give the correct $\langle p_\parallel \rangle$ and $\langle p_\perp \rangle$,
 as well as $\Delta p_\parallel/ p_\parallel$, $\Delta \rho / \rho$
 (the change of these profiles on a flux surface), and anisotropy $\Delta$,
 which are not determined by a choice of the shape of the distribution function \cite{Iacono1990}.
 Here, $\langle ... \rangle$ means flux surface average.
 We also mention that our stability treatment later on does not rely on the choice of equilibrium distribution function,
 as long as the modified GSE is solved self-consistently,
 and can provide $\Psi$ as a function of $(R,Z)$, i.e. the flux surfaces, for the stability treatment.

The solution $\Psi(R,Z)$ for the modified GSE is then mapped into the straight field line coordinates $(s, \vartheta, \varphi)$,
with $s = \sqrt{\Psi}$ and $\vartheta$ defined by
\begin{equation}
\vartheta \equiv \int_\Psi \frac{B_{\varphi} dl} {q RB_p}, \quad
 q(\Psi) \equiv \frac{1}{2\pi}\oint_\Psi \frac{B_{\varphi} dl} {RB_p},
\end{equation}
in which $B_p$ is the poloidal field and $q$ the safety factor.
The integrals are performed on a constant $\Psi$ surface clockwise facing the direction of $\bm{e}_\varphi$ and starting from $Z=Z_0$,
in which $Z_0$ is the $Z$ coordinate for the magnetic axis.
The metric coefficients of this curvilinear coordinate, $g_{ij}$ and $g^{ij}$, as well as the Jacobian $J$,
are defined by
\begin{eqnarray}
 g_{ij} \equiv \nabla x^i \cdot \nabla x^j, \quad
 g^{ij} \equiv \frac{\partial \bm{r}}{\partial x^i} \cdot \frac{\partial \bm{r}}{\partial x^j}, \\
 J \equiv \sqrt{det(g_{ij})} = \frac{f q R}{F},
\end{eqnarray}
where $f = d\Psi/ds$ and $det$ is the determinant operator, 
with$(x^1,x^2,x^3) = (s, \vartheta, \varphi)$.
In the straight field line coordinates, the contravariant equilibrium current is given by
\begin{eqnarray}
 j^1 =  \frac{1}{ J} \frac{\partial F}{\partial \vartheta} , \ \ 
 j^2 = -\frac{1}{ J} \frac{\partial F}{\partial s}, \nonumber\\
 j^3 =  \frac{1}{ J} \left( \frac{\partial}{\partial s} \frac{g_{22}F}{qR^2} 
 - \frac{\partial}{\partial \vartheta} \frac{g_{12}F}{qR^2} \right),
\end{eqnarray}
and the contravariant magnetic field components are given by
\begin{eqnarray}
 B^1 =0, \quad
 B^2 = \frac{F}{qR^2}, \quad
 B^3 = \frac{F}{R^2}.
\end{eqnarray}
For the GSE with anisotropy in the straight field line coordinates, 
one can refer to \rcite{Fitzgerald2015}, as we will not restate it here.

\section{The perturbed equations in the straight field line coordinates} \label{sec:perturbed_eqs}
In this section, we write our first order perturbed equations in the straight field line coordinates using contravariant and/or covariant representatives.
Same as the original MISHKA, a set of ``optimized'' projections of $\tilde{\bm{V}}$ and $\tilde{\bm{B}}$ is used
instead of the contra/co-variant projections.
We use circumflexes to label these projections in order to distinguish them from the contra/co-variant projections,
which are labeled by tildas.
The perturbed fluid velocity $\tilde{\bm{V}}$ is expressed in its contravariant normal component $\tilde{V}^1$,
its binormal projection $\hat{V}^2$ and its parallel projection $\hat{V}^3$, with
\begin{eqnarray}
 \hat{V}^2 = [\tilde{\bm{V}} \times \bm{B}]_1, \quad
 \hat{V}^3 = \frac{\tilde{\bm{V}} \cdot \bm{B}}{B^2}.
\end{eqnarray}
The perturbed magnetic field $\tilde{\bm{B}}$ is calculated by
taking the curl of the perturbed magnetic vector potential $\tilde{\bm{A}}$
(i.e. $\tilde{\bm{B}} = \nabla \times \tilde{\bm{A}}$).
Then similarly, $\tilde{\bm{A}}$ is expressed in its covariant normal component $\tilde{A}_1$,
its binormal projection $\hat{A}_2$ and its parallel projection $\hat{A}_3$, with
\begin{eqnarray}
 \hat{A}_2 = \frac{[\tilde{\bm{A}} \times \bm{B}]^1}{B^2}, \quad
 \hat{A}_3 = \frac{\tilde{\bm{A}} \cdot \bm{B}}{B^2}.
\end{eqnarray}
The conversion between these projections and contra/covariant components of both $\tilde{\bm{V}}$ and $\tilde{\bm{A}}$
can be found in \rcite{Mikhailovskii1997} and \cite{Fitzgerald2015},
while $\tilde{B}^1$, $\tilde{B}^2$ and $\tilde{B}^3$ are related to $\tilde{A}_1$, $\hat{A}_2$ and $\hat{A}_3$ through
Eq. (90) to (92) in \rcite{Fitzgerald2015}.
The covariant components are related to contravariant components through
$\tilde{B}_i = \sum_j g_{ij} \tilde{B}^j$.
Finally, the perturbed magnetic field strength is given by $\tilde{B} = \tilde{\bm{B}} \cdot \bm{b}$.

\subsection{The ideal Ohm's law}
\Eqref{eq:maxwell_1}, the ideal Ohm's law, stays unchanged moving from isotropic plasma to anisotropic plasma.
The equations are therefore identical to \rcite{Mikhailovskii1997}:
\begin{eqnarray}
 \lambda \tilde{A}_1 = \hat{V}^2,  \label{eq:sfl_ohm1}\\
 \lambda \hat{A}_2   =-\tilde{V}_1, \label{eq:sfl_ohm2}\\
 \lambda \hat{A}_3   = 0.
\end{eqnarray}
We recall that $\lambda$ = $\gamma - i \omega$.
When plasma equilibrium flow and resistivity are ignored,
$\hat{A}_3$ is an ignorable component, henceforth neglected.

\subsection{The momentum equation}
Perturbing \eqref{eq:gcp_momentum}, one obtains
\begin{eqnarray}
 \rho_0 \frac{\partial \tilde{\bm{V}}}{\partial t} =  - \nabla \cdot \tilde{\ten{P}} + \bm{H},
 \label{eq:perturbed_momentum}
\end{eqnarray}
in which
\begin{equation}
 \tilde{\ten{P}} = \tilde{\pper} (\ten{I} - \bm{b}\bm{b}) + \tilde{\ppar} \bm{b}\bm{b} 
 + (\ppar - \pper) \left(\frac{\tilde{\bm{B}}_\perp}{B} \bm{b} + \bm{b} \frac{\tilde{\bm{B}}_\perp}{B} \right),
\end{equation}
and
\begin{equation}
 \bm{H} = (\nabla \times \bm{B}) \times \tilde{\bm{B}} - \bm{B} \times (\nabla \times \tilde{\bm{B}}).
\end{equation}
The first two covariant components of $\bm{H}$, $H_1$ and $H_2$, are provided in \rcite{Fitzgerald2015} and restated in \ref{app:auxiliary_formulars}
while $H_3$ is given in \ref{app:auxiliary_formulars} as well.

After some algebra, 
we reach the perturbed momentum equation covariantly in the straight field line coordinates:
\begin{eqnarray}
\fl \lambda \rho \tilde{V}_1 = (1-\Delta) H_1 
 -  \partial_s \tilde{\pper} 
 - \partial_j (\tilde{\ppar} - \tilde{\pper} - 2\Delta B \tilde{B}) \frac{B_1 B^j}{|B|^2}   \nonumber\\
 -\Delta \partial_s (B \tilde{B})
 - (B^j \tilde{B}_1 + \tilde{B}^j B_1) \partial_j \Delta  \nonumber\\
 - (\tilde{\ppar} - \tilde{\pper} - 2\Delta B \tilde{B}) \left(\frac{B_1}{B} \nabla \cdot \bm{b}  + \kappa_1\right), 
 \label{eq:sfl_momentum1} \\
 \fl \lambda \rho \tilde{V}_2 = (1-\Delta) H_2 
 -  \partial_\vartheta \tilde{\pper} 
 - \partial_j (\tilde{\ppar} - \tilde{\pper} - 2\Delta B \tilde{B}) \frac{B_2 B^j}{|B|^2}   \nonumber\\
 -\Delta \partial_\vartheta (B \tilde{B})
 - (B^j \tilde{B}_2 + \tilde{B}^j B_2) \partial_j \Delta  \nonumber\\
 - (\tilde{\ppar} - \tilde{\pper} - 2\Delta B \tilde{B}) \left(\frac{B_2}{B} \nabla \cdot \bm{b}  + \kappa_2\right),
 \label{eq:sfl_momentum2}
\end{eqnarray}
summing over index $j=1,2,3$, in which $\bm{\kappa} = \bm{b} \cdot \nabla \bm{b}$ is the magnetic field line curvature
with its covariant components $\kappa_1$ and $\kappa_2$ given in \ref{app:auxiliary_formulars}.
Taking the dot product of \eqref{eq:perturbed_momentum} with $\bm{B}$, 
the third component of the momentum equation is written as
\begin{eqnarray}
\fl \lambda \rho |B|^2 \hat{V}^3 = 
(1-\Delta)B^j H_i - B^j \partial_j (\tilde{\ppar} - 2\Delta B \tilde{B})
-\Delta B^j \partial_j (B \tilde{B}) \nonumber\\
-\partial_j \Delta (\tilde{B}^j |B|^2 + B^j B \tilde{B})
- (\tilde{\ppar} - \tilde{\pper} - 2\Delta B \tilde{B}) (B \nabla \cdot \bm{b}), 
\label{eq:sfl_momentum3}
\end{eqnarray}
summing over index $j=1,2,3$.

\subsection{The fluid closure equation}
For the single-adiabatic and double-adiabatic model,
the fluid closure equations have similar forms in the straight field line coordinates,
which are given by
\begin{eqnarray}
\fl \lambda \tilde{\ppar} = -\frac{\gparone}{J} [\partial_s(J\tilde{V}^1)+\partial_\vartheta(J\tilde{V}^2)+\partial_\varphi(J\tilde{V}^3)] - \gpartwo E \nonumber\\ 
 - (\tilde{V}^1 \partial_s + \tilde{V}^2 \partial_\vartheta)f_\parallel, \label{eq:sfl_pressure1} \\
 \fl \lambda \tilde{\pper} = -\frac{\gperone}{J} [\partial_s(J\tilde{V}^1)+\partial_\vartheta(J\tilde{V}^2)+\partial_\varphi(J\tilde{V}^3)] - \gpertwo E \nonumber\\ 
 - (\tilde{V}^1 \partial_s + \tilde{V}^2 \partial_\vartheta)f_\perp, \label{eq:sfl_pressure2}
\end{eqnarray}
where
\begin{equation}
 E = \frac{B^j}{B} \partial_j (B\hat{V}^3) 
 - \tilde{V}^1 \kappa_1 - \frac{\hat{V}^2}{fq} \kappa_2.
\end{equation}
For single-adiabatic model, we have
\begin{eqnarray}
 \gparone = \gperone =   \frac{1}{3} \ppar + \frac{4}{3} \pper, \quad
 \gpartwo = \gpertwo =   \frac{2}{3} \ppar - \frac{2}{3} \pper, \nonumber\\
 f_\parallel = f_\perp = \frac{1}{3} \ppar + \frac{2}{3} \pper. \label{eq:gamma_SA}
\end{eqnarray}
For double-adiabatic model, we have
\begin{eqnarray}
 \gparone =  \ppar, \quad
 \gpartwo = 2 \ppar, \quad
 f_\parallel = \ppar,  \nonumber\\
 \gperone = 2 \pper, \quad
 \gpartwo = -  \pper, \quad
 f_\parallel = \pper. \label{eq:gamma_DA}
\end{eqnarray}

There is no need to restate the incompressible fluid closure here,
since \eqref{eq:pressure_IC1} and \eref{eq:pressure_IC2} are already given in the straight field line coordinates.

\section{Numerical method} \label{sec:numerical}
Similar to the original MISHKA and its extension MISHKA-D/F,
we use the following variables in our anisotropic extension of the MISHKA code,
namely MISHKA-A (anisotropy):
\begin{eqnarray}
 X_1 = fq \tilde{V}^1, \ \ X_2 = i \hat{V}^2,  \ \ X_3 = i \tilde{A}_1,  \ \ X_4 = fq\hat{A}_2,\nonumber\\
 X_5 = i f \hat{V}^3,  \ \ X_6 = f \tilde{\pper}, \ \ X_7 = f \tilde{\ppar}.
\end{eqnarray}
These variables are then expanded poloidally and toroidally in Fourier harmonies with mode number $m$ and $n$ respectively,
and radially in cubic/quadratic Hermite elements, i.e.
\begin{equation}
 X_\alpha =e^{\lambda t + in\varphi} \sum_{m=-\infty}^{\infty} \sum_{\nu=1}^{N} X_\alpha^{m \nu} H_\nu(s) e^{im\vartheta },
\end{equation}
in which $H_\nu(s)$ is the cubic/quadratic Hermite elements and $N$ the number of radial elements.
The weak form is constructed by multiplying \eqref{eq:sfl_momentum1}, \eref{eq:sfl_momentum2} , \eref{eq:sfl_momentum3},
\eref{eq:sfl_ohm1}, \eref{eq:sfl_ohm2}, \eref{eq:sfl_pressure1} and \eref{eq:sfl_pressure2} respectively by
$\tilde{V}^{1*}/(1-\Delta)$, $\hat{V}^{2*}/fq(1-\Delta)$, $f\hat{V}^{3*}/(1-\Delta)$, $A_1^*/J$, $f^2 q^2 \hat{A}_2^*/J$, $f\tilde{\ppar}$ and $f\tilde{\pper}$, 
converting the system into a linear algebra problem solving
\begin{equation}
 \lambda N_i = M_i,
\end{equation}
in which
\begin{equation}
 N_i = \sum_{j=1}^{8} \int B(i,j) {X}_i^* X_j J ds d\vartheta,
\end{equation}
and
\begin{eqnarray}
\fl M_i = \sum_{j=1}^{8} \int [A(i,j) {X}_i^* X_j
 + A(i',j)  \frac{\partial {X}_i^*}{\partial s} X_j  \nonumber\\
 + A(i,j')  {X}_i^* \frac{\partial X_j}{\partial s} 
 + A(i',j') \frac{\partial{X}_i^*}{\partial s} \frac{\partial X_j}{\partial s} ]
 J ds d\vartheta.
\end{eqnarray}
We separate the matrix elements $A(i, j)$ into 
\begin{equation}
 A(i, j) = A^0(i, j) + A^A(i, j),
 \label{eq:amat}
\end{equation}
in which $A^0(i, j)$ are the common terms for MISHKA (isotropic) and MISHKA-A (anisotropic)
and $A^A(i, j)$ are terms existing only in anisotropic plasmas.
These matrix elements are given in \ref{app:matrix_elements}.

To obtain the continuous spectrum,
we reduced MISHIKA-A to a continuum code (CSMISH-A). 
The method provided in Poedts \etal \cite{Poedts1993} (CSCAS) is implemented here,
carrying the calculation in the vicinity of the singularity $\Psi \rightarrow \Psi_0$.

\section{Anisotropy impact on plasma continuous spectrum} \label{sec:continuum}
In this section, we study the continuous spectrum of an anisotropic plasma described by the SA and the CGL model,
as well as the modification of anisotropy to the continuous spectrum.
We present a set of examples with circular cross-section,
large aspect ratio ($\epsilon = 0.3$) and low $\beta$.
The equilibrium solutions are computed by HELENA+ATF \cite{Qu2014} using the enthalpy form of the modified GSE with the bi-Maxwellian distribution 
and the equilibrium thermal closure $T_\parallel = T_\parallel(\Psi)$. 
We start from an isotropic MHD reference case with
\begin{eqnarray}
  T(\bPsi) = C_0(1-\bPsi)^2 + C_1, \quad
  RB_\varphi(\bPsi) = F_{0}, \nonumber\\
  H(\bPsi) = \frac{C_0}{2}(1-\bPsi)^3 + C_2, 
  \label{eq:profile-choice}
\end{eqnarray}
where $\bPsi$ is the normalized flux surface defined as $\bPsi=0$ on axis and $\bPsi=1$ at the edge,
and $C_0, C_1, C_2$ and $F_0$ are adjustable constants.
Constant $F_{0}$ indicates vacuum field strength.
Constants $C_1$ and $C_2$ are small values to make density and current profiles vanish at the plasma edge.
The density and pressure profiles are given by \eqref{eq:rho_tper} and \eref{eq:ppar_pper}.
The $q$ profile monotonically increases from $q_0 = 1.7$ to $q_{95} = 7$.
We choose $\beta = 1\%$ on the magnetic axis.
In the next step, we add anisotropy to this reference equilibrium.
The $\Theta$ profile, which indicates the magnitude of anisotropy,
is chosen to be constant.
Therefore, anisotropy decreases from core to edge following the same trend of $T$,
which is associated with on-axis beam heating or ICRH.
For an individual anisotropic equilibrium,
we specify a $\Theta_0$, then iterate the $T_\parallel$, $F_M$ and $H$ profiles to keep
$\langle p^* \rangle$, $\langle j \rangle$ and $\langle \rho \rangle$
on each flux surface identical to the isotopic reference case.
Here $p^* = (\ppar + \pper)/2$ and $\langle ... \rangle$ means flux surface average.
In this way the $q$ profile and the metrics of these anisotropic equilibria are the same as the reference isotropic case to $O( \epsilon^2 (\ppar-\pper)/\ppar)$.
We have accordingly obtained equilibria ranging from $\pper = 1.7 \ppar$ (perpendicular beam or ICRH) to $\ppar = 1.8\pper$ (parallel beam) at core.
 When we go to higher anisotropy like $\pper > 1.7 \ppar$ and $\ppar > 1.8\pper$,
 we are unable to reduce the difference of $q_0$ between an anisotropic case (for example $\pper = 2 \ppar$) 
 and its opposite case ($\ppar = 2 \pper$) to less than $1\%$ when we fix other parameters,
 since the flux surfaces of an anisotropic equilibrium is not completely reproducible by an isotropic equilibrium, 
 or an anisotropic equilibrium with opposite magnitude of anisotropy \cite{Qu2014}.
 Our start point is to identify the difference of anisotropic stability with equilibria in almost same conditions.
 Consequently, these higher anisotropy regimes are not explored here,
 because we are unable to keep them in these same conditions.
 However, our model and code are capable to describe cases with higher anisotropy,
 such as the $\pper = 2.5 \ppar$ discharge in JET. 

\begin{figure}[!htbp]
  \centering
  $\begin{array}{c c}
   \includegraphics[width=6cm]{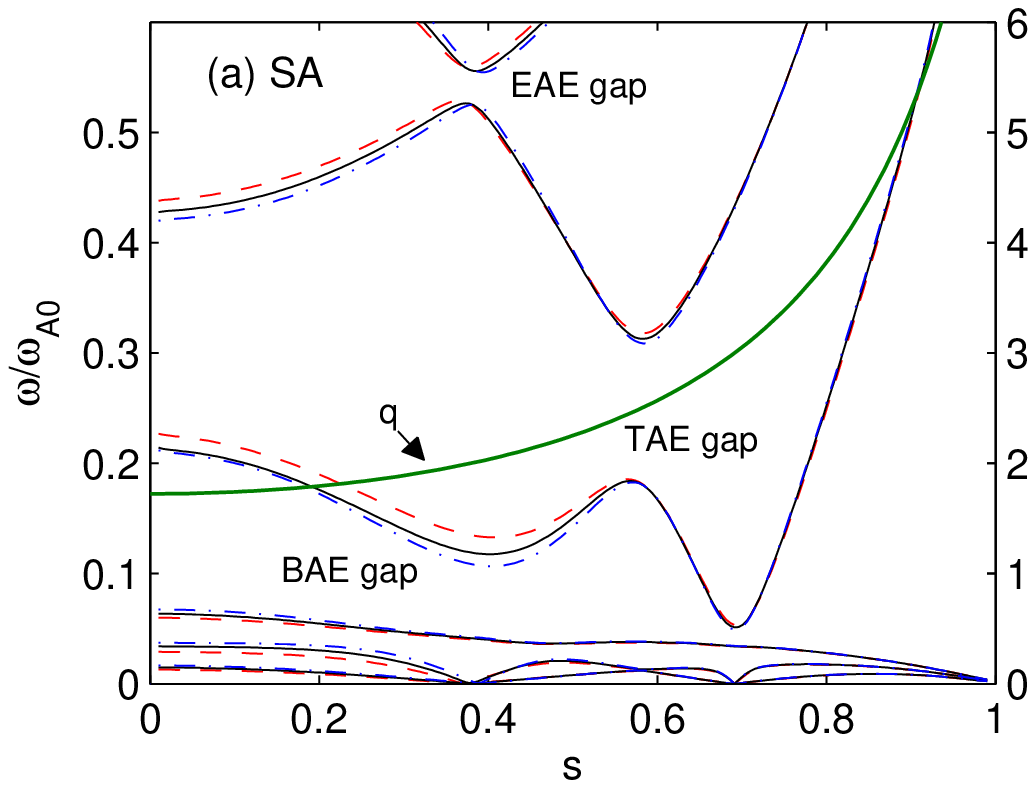} & \includegraphics[width=6cm]{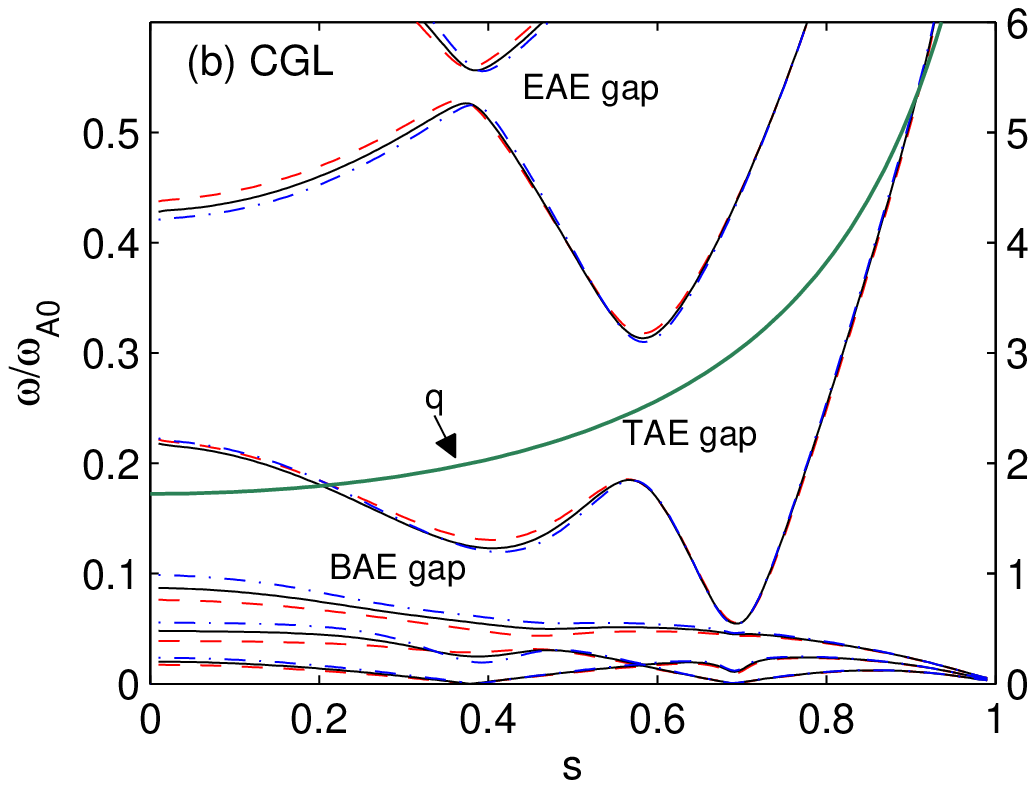}
  \end{array}$
  \caption{The $n=-1$, $m=1,2,3$ continuous spectrum (left axis) and the $q$ profile (right axis) of a plasma with (a) SA closure (b) CGL closure.
  The frequency is normalized to $\omega_{A0}$, the \Alfven frequency on axis,
  and $s = \sqrt{\bPsi}$ is the standard flux label. 
  The red dash line, black solid line and blue dash dot line shows respectively the cases with
  $\pper = 1.7 \ppar$, $\pper = \ppar$ and $\ppar = 1.8\pper$ on axis.
  The EAE gap, TAE gap and BAE gap are labeled in each figure.}
  \label{fig:continuum_compare}
\end{figure}

The continuous spectrum of these examples are then computed by CSMISH-A.
\Figref{fig:continuum_compare} shows the $n=-1$ and $m=1,2,3$ continuous spectrum of three cases :
$\pper = 1.7 \ppar$, $\pper = \ppar$ and $\ppar = 1.8\pper$ on axis, for (a) the SA model and (b) the CGL model.
The linear growth rate of the continuous spectra in all these examples is observed to have $\gamma < 10^{-8} \omega_A$.
We note that the small growth rate here is due to numerical errors (e.g. finite grid resolution) 
and is reduced by improving numerical precision.
Therefore, we conclude that these continuous modes are stable.
As in the ideal MHD spectrum, 
two sets of branches, a shear \Alfven set ($\omega/\omega_{A0}>0.1$) and a slow sound set ($\omega/\omega_{A0}<0.1$),
appear at higher frequency and lower frequency, respectively.
A resonance between $m=2$ and $m=3$ shear \Alfven branches 
occurs at $q=2.5$ surface and forms the TAE gap ($\Delta m = 1$ gap) around $s=0.6$.
Meanwhile, a resonance between $m=1$ and $m=3$ forms the EAE gap ($\Delta m = 2$ gap) at $q=2$ surface around $s=0.4$.
The coupling between the shear \Alfven and the slow branches forms the low frequency gaps ($\Delta m = 0$ BAE gap).
Moving to the edge, frequencies of the shear \Alfven branches approach infinity as density approaches zero,
while frequencies of the slow waves vanish as pressure goes to zero.

\Figref{fig:continuum_compare} also demonstrates the modification of anisotropy to the continuous spectrum.
Anisotropy does not modify the main structure of the spectrum and the position of the gaps,
but shifts the gaps and branches.
For both models, around the core where the magnitude of anisotropy is higher,
the difference between the three cases with different anisotropy is more significant.
At the edge where anisotropy is vanishing, the three spectra merge to one.
For the $\ppar = 1.8\pper$ case described by the SA model,
all the shear \Alfven branches are lowered ($0.01 \omega_{A}$ on axis),
while the slow branches are shifted up ($7\%$ on axis).
For the CGL model, the lowest shear Alfven branch is almost unchanged,
while the frequency of the slow branches increases by $14\%$ on axis.
The modification to slow branches will be investigated in \secref{sec:cylindrical_limit}.
The change of the shear \Alfven branches can be explained by 
the change of these branches' coupling to plasma compressibility through geodesic curvature,
with different anisotropy and different model.
Also, the $q$ profile is only conserved to the reference isotropic case to $O(\epsilon^2 (\ppar-\pper)/\ppar)$.
With $\epsilon = 0.3$ in our example, the change of $q_0$ is $0.01$ (of $1.7$) for the $\ppar = 1.8\pper$ compared to the isotropic reference case,
which will sightly modify all the branches.
Looking at continuum gaps, the upper and lower accumulation points of both the TAE gap and the EAE gap are almost unchanged,
meanwhile the upper accumulation point of the BAE gap is shifted up for both models ($8\%$ for SA and $4\%$ for CGL).
For the $\pper = 1.7\ppar$ case, all the above modifications are reversed,
with a similar magnitude of change.

To understand the modification of anisotropy and the above differences,
we study two specific feature of the continuous spectrum: its cylindrical limit and the low frequency BAE gap.
The former one determines the main frequency of both the shear Alfven and the slow branches,
and the latter describes the shear \Alfven and slow coupling.

\subsection{the Cylindrical limit} \label{sec:cylindrical_limit}

In the cylindrical limit, the equilibrium quantities are free of poloidal angle dependency.
Therefore the coupling between two shear \Alfven branches vanishes.
Also, the geodesic curvature, which couples the shear \Alfven branches and the slow branches, is zero.
Building on \rcite{Fitzgerald2015}, we have computed the continuum in the cylindrical limit.
We retain the ignored $(\ppar - \pper)(\bm{b}_1 \bm{b} + \bm{b} \bm{b}_1)$ term in the perturbed pressure tensor in \rcite{Fitzgerald2015},
therefore the missing firehose factor $1-\Delta$ for the single-adiabatic \Alfven branches is now recovered.
The frequency of mode $(m,n)$ is now simply given by
\begin{eqnarray}
 \omega_{A,SA}^2=\omega_{A,CGL}^2 = \frac{(1-\Delta)B^2}{\rho R_0^2} (m/q + n)^2, \label{eq:continuum_cyl_a} \\ 
 \omega_{S,SA}^2 = \frac{ \ppar + \frac{2}{3} \pper }{\frac{1}{3} \ppar + \frac{4}{3} \pper  + B^2}
 \frac{B^2}{\rho R_0^2} (m/q + n)^2, \label{eq:continuum_cyl_s_SA} \\
 \omega_{S,CGL}^2 = \frac{3 \ppar }{2 \pper  + B^2}\frac{B^2}{\rho R_0^2} (m/q + n)^2, \label{eq:continuum_cyl_s_CGL}
\end{eqnarray}
where $R_0$ is the major radius of the magnetic axis.
Here, ``A'' in the subscript labels the shear \Alfven branches and ``S'' labels the slow branches.
Inspection of \eqref{eq:continuum_cyl_a} shows that the cylindrical shear \Alfven continuum is not fluid closure dependent.
The anisotropy modifies these branches by the firehose factor $1-\Delta$.
This is consistent with previous results \cite{Kato1966, Hellsten1984}.
In contrast, the slow branches, as shown by \figref{fig:continuum_compare}, 
have strong fluid closure dependency and anisotropy dependency,
with $\omega_{S,SA} \neq \omega_{S,CGL}$ even when the equilibrium is isotropic.
In the isotropic limit, the SA model reduces to the result given by ideal MHD with adiabatic gas law,
while the CGL model does not converge to ideal MHD.
Indeed, the frequency $\omega_{S,CGL}$ is roughly $35\%$ larger than $\omega_{S,SA}$ when the plasma is isotropic.
As in \eqref{eq:continuum_cyl_s_SA} and \eref{eq:continuum_cyl_s_CGL}, 
the frequency of the slow branches with both model are increasing when $\ppar/\pper$ increases,
if $\langle p^* \rangle$ is kept constant,
although CGL model shows more significant change compared to SA. 
We have compared the result from CSMISH-A in the cylindrical limit (very large aspect ratio) 
with \eqref{eq:continuum_cyl_a} to \eref{eq:continuum_cyl_s_CGL} for both SA and CGL,
showing very good agreement.

\subsection{The BAE gap change due to anisotropy}
\begin{figure}[!htbp]
  \centering
  $\begin{array}{c c}
   \includegraphics[width=6cm]{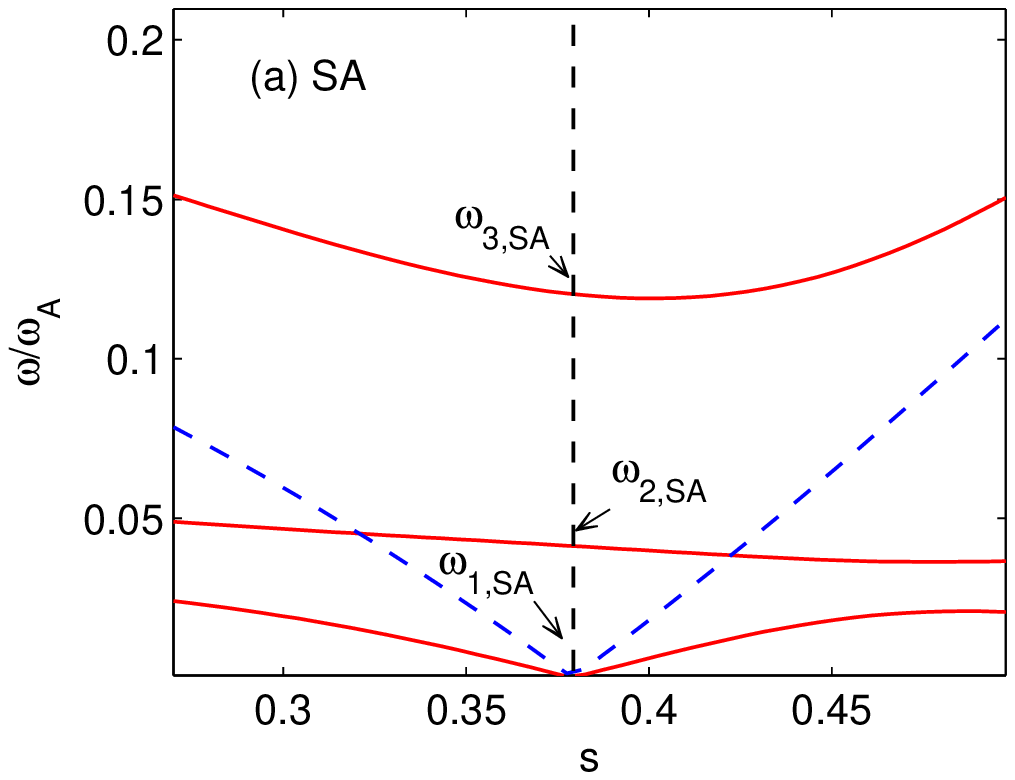} & \includegraphics[width=6cm]{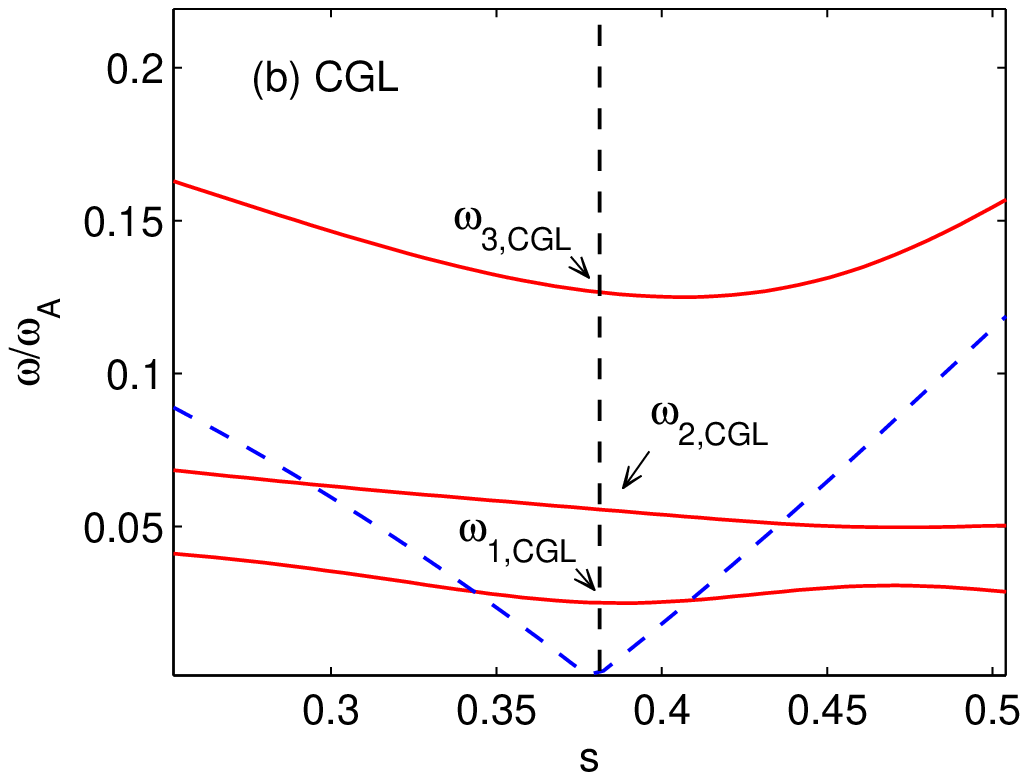}
  \end{array}$
  \caption{Zooming into the $q=2$ BAE gap of the $n=-1$ continuous spectrum 
  of the anisotropic case in \figref{fig:continuum_compare} with $\ppar = 1.8\pper$ on axis,
    for (a) the SA model and (b) the CGL model.
    The blue dash lines are the incompressible $m=2$ shear \Alfven branch.
    The vertical lines indicate the flux surface where $q=2$, and the incompressible $m=2$ shear \Alfven branch hits zero.
    The red solid lines are the coupled $m=2$ shear \Alfven branch and $m=1,3$ slow branches due to finite compressibility (SA or CGL), 
    with the frequency of the upper, middle and lowest branches labeled as $\omega_3$, $\omega_2$ and $\omega_1$ at $q=2$ surface, respectively. }
  \label{fig:bae_gap_zoom}
\end{figure}

The low frequency gap (BAE gap)\cite{Turnbull1993} appears on the resonant flux surface where $m+nq=0$, 
and is induced by the finite compressibility of the plasma.
Inspection of \figref{fig:continuum_compare} shows that for different magnitude of anisotropy,
the width of this gap is changed.
Also, the gap width is different for the SA and the CGL model,
implying its dependency on fluid closure model.
\Figref{fig:bae_gap_zoom} zooms in into the $q=2$ BAE gap in \figref{fig:continuum_compare} 
for the anisotropic case with $\ppar = 1.8\pper$ on axis.
Only the major $m=2$ shear \Alfven branch and the $m=2 \pm 1$ slow side bands are shown here.
In \figref{fig:bae_gap_zoom}, the frequency of the upper, middle and lowest branches on the resonant flux surface (located at $s=0.38$) are labeled
as $\omega_3$, $\omega_2$ and $\omega_1$ respectively.
The BAE gap of the SA model has the same structure as an isotropic plasma described by the MHD model.
Its lowest branch approaches zero when $m+nq=0$, i.e. $\omega_1 = 0$.
To the contrary, in CGL we have $\omega_1 > 0$, inducing an additional gap at very low frequency.

\begin{figure}[!htbp]
  \centering
  $\begin{array}{c c}
   \includegraphics[width=6cm]{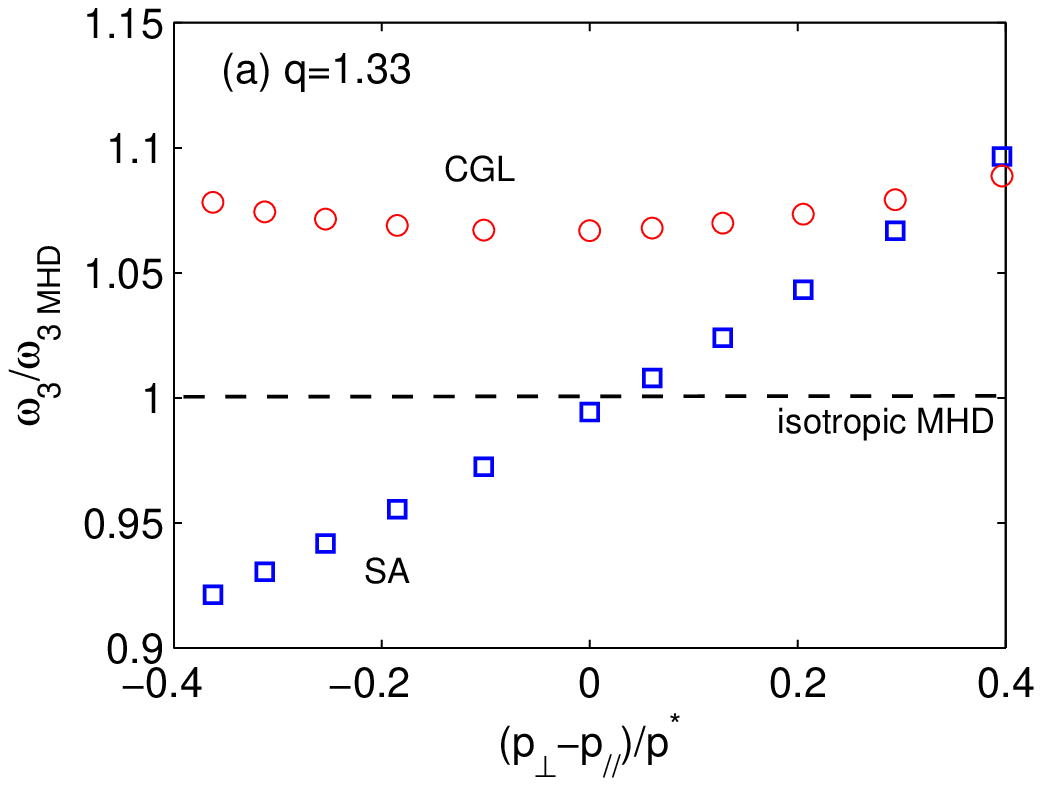} & \includegraphics[width=6cm]{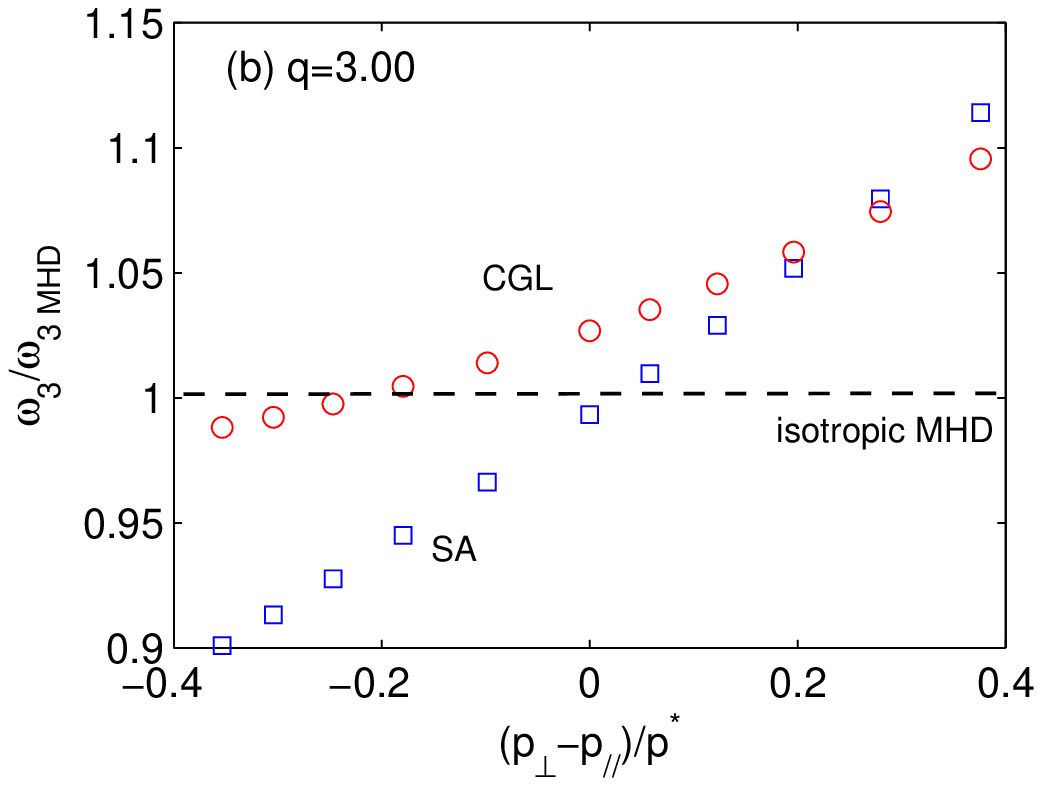}
  \end{array}$
  \caption{The change of the BAE gap upper accumulation point frequency ($\omega_3$) due to the change of anisotropy for
           a BAE gap with (a) $q=1.33$, $n=-3$ (b) $q=3.00$, $n=-1$.
	   The local magnitude of anisotropy is described by the relative difference of $\ppar$ and $\pper$, i.e. $(\ppar-\pper)/p^*$.
	   The frequency of $\omega_3$ is normalized to the analytic ideal MHD value of $\omega_3$ for the reference isotropic case, 
	   as shown by the horizontal dash line.
	  The symbols are numerical results from the CSMISH-A code:
	  blue squares and solid lines for SA, red circles and solid lines for CGL.
	   }
  \label{fig:bae_gap_change}
\end{figure}

In this section, we are only interested in $\omega_3$, the upper accumulation point of a BAE gap,
which determines the gap width.
We study two separate cases, with the gap located at a low $q$ position ($q=1.33$) and a high $q$ position ($q=3$),
as shown in \figref{fig:bae_gap_change} (a) and (b), respectively.
The frequencies in \figref{fig:bae_gap_change} are normalized to the analytic ideal MHD value of $\omega_3$ 
for the reference isotropic case \cite{Winsor1968}, written as
\begin{equation}
 \omega^2_{3,MHD} = \frac{2 \gamma p} {(\gamma p + B^2)\rho R_0^2} \left(1 + \frac{1}{2q^2} \right),
\end{equation}
with $\gamma = 5/3$.
\Figref{fig:bae_gap_change} (a) shows that for $q=1.33$, 
the SA closure gives a greater $\omega_3$ when $\ppar > \pper$, and a smaller $\omega_3$ when $\ppar < \pper$.
It's almost a linear function of $(\ppar-\pper)/p^*$.
The change of $\omega_3$ is roughly $8\%$ for $\ppar \approx 1.5 \pper$ or $\pper \approx 1.5 \ppar$,
the farthest right and left data points in the figure.
For the CGL closure, $\omega_3$ is $7\%$ higher than the isotropic ideal MHD reference case.
It's dependency on $(\ppar-\pper)/p^*$ is almost negligible.
Moving to \figref{fig:bae_gap_change} (b) where $q=3$,
in SA model the dependency of $\omega_3$ on $(\ppar-\pper)/p^*$ becomes higher,
with a $12\%$ change for $\ppar \approx 1.5 \pper$ or $\pper \approx 1.5 \ppar$.
Meanwhile, the ratio $\omega_3/\omega_{3MHD}$ decreases to $1.03$ in the isotropic case,
and the $\omega_3$ for CGL has a weak dependency on anisotropy:
about a further $5\%$ change for the extreme cases.

\section{Anisotropy impact on the internal kink mode} \label{sec:internal_kink}
In this section, we study the impact of anisotropy on the $n=1$ internal kink mode 
in a tokamak plasma with large aspect ratio ($\epsilon$ = 0.1) and circular cross section.
This also serves as a benchmark of MISHKA-A working as a global normal mode code.
For simplicity, the equilibrium distribution function is taken to be bi-Maxwellian.

We start from a reference isotropic equilibrium with the current profile and the pressure profile taking the form,
\begin{eqnarray}
  \langle j \rangle = j_0 (1 - \bPsi), \\
  \ppar = \pper = p_0(1 - \bPsi),
\end{eqnarray}
where $j_0$ and $p_0$ are constants.
The density profile is taken to be constant, i.e. $\rho = \rho_0$.
The safety factor on axis, $q_0$, and the ratio of kinetic energy to magnetic energy, $\beta$, 
can then be adjusted by changing the ratio $p_0/j_0$ and the vacuum field.
The safety factor $q$ is monotonically increasing: only one $q=1$ surface exists in the plasma.
Similar to \secref{sec:continuum},
based on this reference isotropic case  we change the $\Theta$ profile with $\Theta = \Theta_0$ in our equilibrium code HELENA+ATF,
meanwhile keeping $\langle p^* \rangle = (\langle \pper \rangle + \langle \ppar \rangle)/2$, $\langle j \rangle$ and 
$\langle \rho \rangle$ unchanged.
In such a way the $q$ profile and metrics are identical to our reference isotropic case to $O(\epsilon^2)$.
The relative anisotropic profile is then approximately given by
\begin{equation}
 \frac{\langle \pper \rangle}{\langle \ppar \rangle} 
 = \frac{1}{1 - \alpha(1-\bPsi)},
\end{equation}
with which the magnitude of anisotropy peaks on axis and vanishes at the boundary.
Here $\alpha$ is an adjustable constant proportional to $\Theta_0$.

In the incompressible limit, the plasma kinetic response to the perturbation is ignored.
The stability of the internal kink mode is determined by the sign
of the perturbed fluid toroidal potential energy $\delta W_{T}$.
When $\delta W_{T} < 0$, the plasma is unstable.
According to the analytical calculation of Bussac \etal \cite{Bussac1975} and Mikhailovskii \cite{Mikhailovskii1983, Mikhailovskii1998book},
the stability criterion of the $n=1$ internal kink in such a scenario, namely the generalized Bussac criterion, is described by 
\begin{equation}
 \delta w + \beta_{pA} > 0,
 \label{eq:generalized_bussac}
\end{equation}
where $\delta w$ is a quadratic function of the value of $\beta_p$ on the $q=1$ surface,
with the coefficients determined by the $q$ profile.
The quantity Bussac $\beta_p$, as a indication of the pressure gradient,
is defined as
\begin{equation}
 \beta_p (\Psi) \equiv \frac{2 [\bar{p}(\Psi) - p(\Psi)]}{B_p^2(\Psi)},
 \label{eq:betap_defination}
\end{equation}
where $\bar{p}$ is the average pressure inside the certain flux surface, i.e.
\begin{equation}
\bar{p}(\Psi_1) \equiv \int_{\Psi<\Psi_1} p dS / \int_{\Psi<\Psi_1} dS.
\end{equation}
For anisotropic plasma,
$\beta_p$ is replaced by $\beta_{p*} \equiv (\beta_{p\parallel} + \beta_{\pper})/2$.
The second term in \eqref{eq:generalized_bussac}, $\beta_{pA}$,
is obtained from \eqref{eq:betap_defination} replacing $p$ by $(\ppar + \pper + \ch)/2$,
and taking the value on the $q=1$ surface as well,
where $\ch$ is defined through partial derivative of $\pper$ as
\begin{equation}
 B \left( \frac{\partial \pper}{\partial B} \right)_\Psi = 2 \pper + \ch.
\end{equation}
For a bi-Maxwellian plasma, $\ch$ is simplified to 
\begin{equation}
 \ch_{bM} = - \frac{2 \pper^2}{\ppar}.
 \label{eq:hatc_bM}
\end{equation}

The generalized Bussac criterion takes into account only the lowest order of the poloidal variation of $\tilde{\ppar}$ and $\tilde{\pper}$,
and neglects the shaping effect \cite{Madden1994} of pressure anisotropy,
leading to its discrepancy from full numerical results when the fast particle distribution function 
has strong and/or complicated poloidal dependency (e.g. with neutral beam heating) \cite{Graves2003}.
The bi-Maxwellian plasma we use here only has a weak poloidal dependency,
satisfying the use of the generalized Bussac criterion.
From \eqref{eq:hatc_bM}, $\beta_{pA}$ is positive when $\ppar > \pper$ and negative when $\ppar < \pper$.
We would expect the plasma to become less stable compared to the reference isotropic case if $\pper > \ppar$ ($\alpha>0$)
and more stable when $\pper < \ppar$ ($\alpha<0$).

\begin{figure}[!htbp]
  \centering
  $\begin{array}{c c}
   \includegraphics[width=6cm]{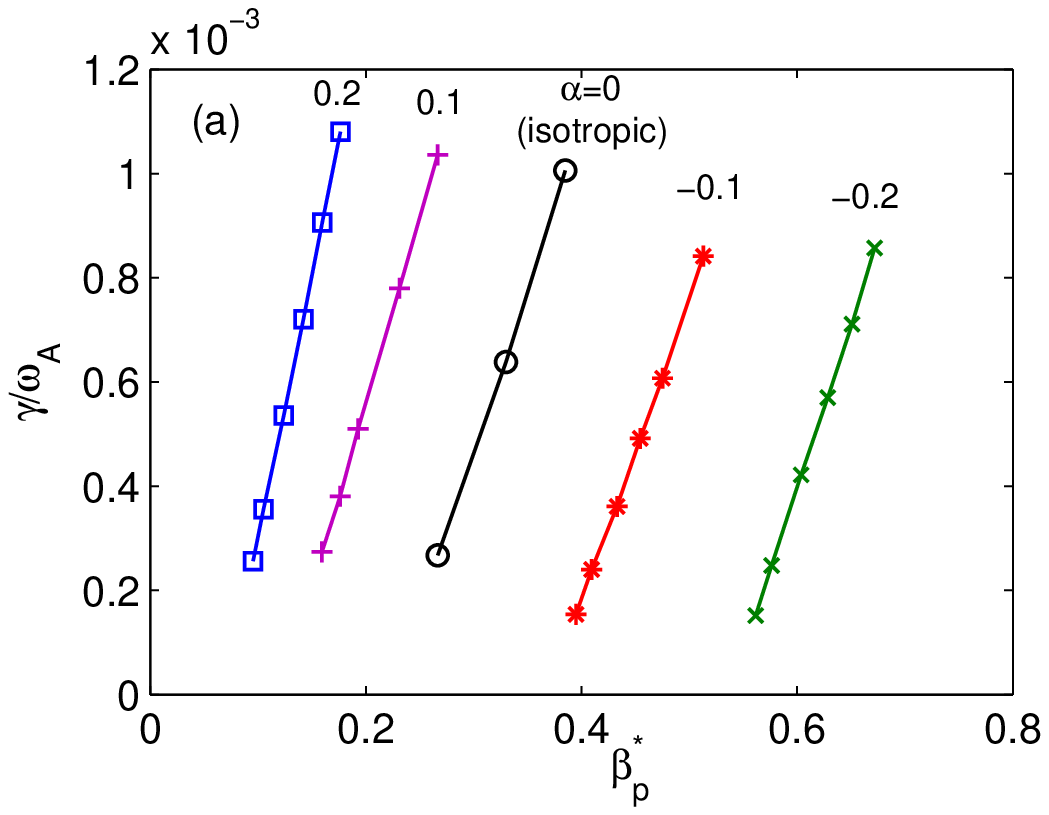} & \includegraphics[width=6cm]{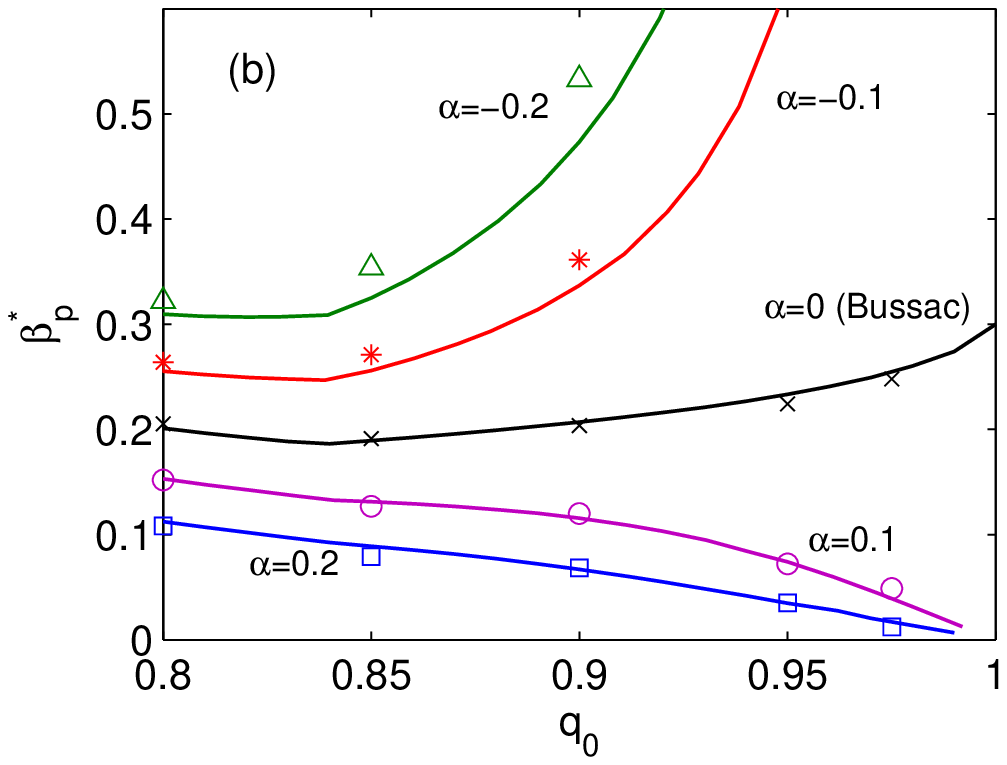}
  \end{array}$
  \caption{(a) The growth rate of the $n=1$ internal kink mode 
  as a function of $\beta_p^{*2}$ for a plasma with $q_0=0.9$.
  The parameter $\alpha$ determines the magnitude of anisotropy,
  with $\pper > \ppar$ for $\alpha>0$ and $\pper < \ppar$ for $\alpha < 0$.
  The growth rate $\gamma$ is normalized to \Alfven velocity $V_A$.
  (b) The modified Bussac critical $\beta_p^{*}$ as a function of $q_0$ for different anisotropy magnitute $\alpha$.
  The lines are analytical result calculated from \eqref{eq:generalized_bussac} and the symbols are numerical results extrapolated from (a).}
  \label{fig:gammavsbetap2}
\end{figure}

To obtain the marginal stability boundary numerically,
we plot the internal kink growth rate as a function of $\beta_p^*$ for different $\alpha$ in \figref{fig:gammavsbetap2} (a).
\Figref{fig:gammavsbetap2} (a) shows that in anisotropic plasma, same as Bussac \etal,
the linear growth rate of the internal kink mode increases with $\beta_p^*$.
For the same $\beta_p^*$, the growth rate is higher when $\alpha$ becomes more positive. 
On the other hand, the growth rate is reduced, or the mode is stabilized, when $\alpha$ becomes more negative.
This is in agreement with the prediction of the generalized Bussac criterion.

The critical $\beta_p^*$ at marginal stability is extrapolated from \figref{fig:gammavsbetap2} (a) 
by fitting $\gamma$ into a quadratic function of $\beta_p^*$ and 
obtaining the fitted curve's intersection with the x axis.
Picking different $q_0$ and different $\alpha$,
the marginal stability boundary is then plotted in \figref{fig:gammavsbetap2} (b)
with a comparison against \eqref{eq:generalized_bussac}.
\Fref{fig:gammavsbetap2} (b) shows that when $\alpha = 0$, i.e. the plasma is isotropic,
the stability limit given by MISHKA-A is in good agreement of the analytical Bussac limit.
When $\alpha > 0$ ($\pper > \ppar$),
the anisotropic incompressible fluid force is destabilizing,
reducing the required pressure gradient to drive the instability.
On the other hand, if $\alpha < 0$ ($\pper < \ppar$),
the anisotropic geometry is stabilizing.
We note that when $q_0$ is close to unity,
the stabilizing/destabilizing effect is greater, 
pushing the stability limit further from the original Bussac limit.
This is due to the fact that when $q_0$ is close to $1$,
the first term in \eqref{eq:generalized_bussac}, $\delta w$, is smaller.
Therefore a tiny change in $\beta_{pA}$ will lead to a dramatic impact of the stability limit.
We also note that the magnitude of anisotropy in \figref{fig:gammavsbetap2} is small 
(with $\pper = 1.25\ppar$ on axis for $\alpha = 0.2$, or $\ppar = 1.2 \pper$ on axis for $\alpha=-0.2$).
We would thus expect that a moderate or large anisotropy will have a much greater impact to the $n=1$ internal kink mode.

We observe a small discrepancy between the generalized Bussac criterion (lines) and
the numerical result (symbols) in \figref{fig:gammavsbetap2}(b) for the $\alpha < 0$ cases.
One possible reason is that in the derivation of the generalized Bussac criterion, 
the eigenfunction is assumed to stay the same as the isotropic reference case.
Also, the perturbed parallel electric field $\tilde{B}$ and the perturbed parallel flow $\tilde{\bm{V}}\cdot \bm{b}$ are ignored.
These neglected features, when taken into account numerically, may have some impact on the marginal stability limit.
Nevertheless, \figref{fig:gammavsbetap2} (b) gives a fairly good benchmark of the MISHKA-A code.

The above treatment ignores the compressional response of the plasma and keeps only the incompressible part.
According to the kinetic theory, the compressional response can either be stabilizing or destabilizing,
depending on the fast particle distribution function, the diamagnetic effects, FLR/FOW effects and other non-ideal effects
(see for example the review of Graves \etal \cite{Graves2005} and Chapman \etal \cite{Chapman2011}).
A full treatment of the $n=1$ internal kink mode will require a $\delta f$ method and possibly the involvement of a kinetic code.
Nevertheless, we can still conclude on that the anisotropic incompressible fluid force of a plasma with $\pper < \ppar$ ($\alpha<0$) is more stable than its isotropic counterpart,
and therefore needs less stabilizing effects from kinetic response to stabilize,
while a plasma with $\pper > \ppar$ ($\alpha>0$) needs more.

Finally, we investigate the compressional response of a plasma described by the CGL model.
We couldn't find any unstable modes for our choice of current and pressure profile,
despite a scan across parameters $0.6<q_0<1$, $0<\beta_p^*<0.5$ and $0.5<\pper/\ppar<2$.
It's long been known that for isotropic plasma we have \cite{Kruskal1958,Rosenbluth1959}
\begin{equation}
 \delta W_{MHD} < \delta W_{K} < \delta W_{CGL},
\end{equation}
where $\delta W_{K}$ is the perturbed potential energy given by the kinetic theory.
For anisotropic plasma, although not rigorously proved, it is very likely to have $\delta W_{K} < \delta W_{CGL}$.
With the CGL gives a prediction that the plasma is stable,
we can conclude that for our choice of profiles and parameter space,
it is possible to stabilize the internal kink mode by plasma compressional response.

\section{Conclusion} \label{sec:conclusion}
We derived and implemented the linearized fluid equations with anisotropy in the straight field line coordinates 
based on three fluid closures: the double-adiabatic model (CGL), the single-adiabatic (SA) model,
and the incompressible model.
The ideal MHD normal mode code MISHKA has then been extended to its anisotropic pressure version, 
MISHKA-A (and the continuous spectrum code, CSMISH-A).
Using these numerical tools, we find that anisotropy mainly modifies the continuous spectrum by changing the slow branches and the BAE gap.
The change of the slow branches is in accordance with the analytical result, 
with a different prediction for the SA model and the CGL model.
For the BAE gap, the lowest branch touches zero at the resonance flux surface for SA/MHD, but does not for CGL.
Meanwhile the change in frequency of the upper accumulation point depends on the local $q$ value, 
the magnitude of anisotropy and the fluid closure.
Finally, we study the impact of anisotropy to the internal kink mode numerically.
If only the incompressible fluid force is considered, 
we find that for a bi-Maxwellian plasma, 
the marginal stability boundary is in good agreement with the analytical result of Bussac \etal and Mikhailovskii:
the plasma is stabilized if $\pper < \ppar$ and destabilized if $\ppar > \pper$.
Also, a parameter scan reveals that for our choice of profiles the internal kink mode is stable,
if the CGL closure is implemented.
This indicates the possibility for these modes to get stabilized by the plasma compressional response,
and that CGL is too strong for the estimation of instabilities.

In this work we restrict our study to large aspect ratio, low beta plasma,
when the equilibrium can be reproduced similarly by an isotopic equilibrium with an $O(\epsilon^2 (\ppar-\pper)/\ppar )$ difference.
In the future, we plan to study the impact of anisotropy on global eigenmodes,
and the possibility of using these eigenmodes as MHD spectroscopy to infer pressure anisotropy.
For example, as indicated by the change of the BAE gap due to anisotropy,
the corresponding modification to a global BAE may serve as an estimation of pressure anisotropy or a validation of the fluid closure model.
We also plan to investigate tokamak plasmas with high $\beta$, low aspect ratio and large anisotropy,
where the current profile and $q$ profile are dramatically modified by anisotropy,
and where the anisotropy shaping effect is important.
Finally, we plan to study experimental data from MAST,
with the equilibria reconstructed anisotropicly by the EFIT-TENSOR code \cite{Fitzgeral2013},
and compute the wave-particle interaction.

\ack
We gratefully acknowledge Dr. Guido Huysmans (ITER) and Dr. Sergei Sharapov (CCFE) for providing the MISHKA and HELENA source code and their permission to use the name MISHKA.
We also thank useful discussion with Em.Prof. Bob Dewar, Dr. Greg von Nessi and Dr. Graham Dennis (ANU) about the physics of fluid closures.
This work is funded by China Scholarship Council, the Australian Institute of Nuclear Science and Engineering (AINSE) Postgraduate
Research Award, Australian ARC project DP1093797 and FT0991899.
This work was part funded by the RCUK Energy Programme [grant number EP/I501045] and by Fusion for Energy. 
To obtain further information on the data and models underlying this paper, 
whose release may be subject to commercial restrictions, please contact \textit{PublicationsManager@ccfe.ac.uk}. 
The views and opinions expressed do not necessarily reflect those of Fusion for Energy which is not liable for any use that may be made of the information contained herein.

\appendix
\section{Auxiliary formulas} \label{app:auxiliary_formulars}
Here we present the formular for $H_1$, $H_2$, $H_3$ and covariant components of the magnetic field line curvature $\bm{\kappa}$:
\begin{eqnarray}
 \fl H_1 = J(j^2 \tilde{B}^3 - j^3 \tilde{B}^2) - \frac{F}{qR^2} \partial_s(g_{12} \tilde{B}^1 + g_{22} \tilde{B}^2) \nonumber\\
 +\frac{F}{qR^2}(\partial_\vartheta + q \partial_\varphi) (g_{11} \tilde{B}^1 + g_{12} \tilde{B}^2) - \frac{F}{R^2} \partial_s(R^2 \tilde{B}^3),\\
 \fl H_2 = J(j^3 \tilde{B}^1 - j^1 \tilde{B}^3) + \frac{F}{R^2} \partial_\varphi (g_{12} \tilde{B}^1 + g_{22} \tilde{B}^2) \nonumber\\
 - \frac{F}{R^2} \partial_s(R^2 \tilde{B}^2), \\
 \fl H_3 = J(j^1 \tilde{B}^2 - j^2 \tilde{B}^1) - B^2 \partial_\varphi(g_{12} \tilde{B}^1 + g_{22} \tilde{B}^2) \nonumber\\
 + B^2 \tilde{B}^3 \partial_\vartheta R^2,
\end{eqnarray}
\begin{eqnarray}
 \fl \kappa_1 =-\frac{F}{q B R^2} \left( \frac{\partial}{\partial s} \frac{q |\nabla \Psi|^2}{B F}  
 + q \frac{\partial}{\partial s} \frac{F}{B} 
 + fq \frac{\partial}{\partial \vartheta} \frac{\nabla \Psi \cdot \nabla \vartheta}{BF} \right), \\
 \fl \kappa_2 =-\frac{F}{R^2 B} \frac{\partial}{\partial \vartheta} \left(\frac{F}{B} \right),\\
 \fl \kappa_3 = -\frac{\kappa_2}{q}.
\end{eqnarray}

\section{Matrix elements} \label{app:matrix_elements}
\subsection{The momentum equation}
The left-hand sides matrix elements $B(1,1)$, $B(1,2)$, $B(2,1)$ and $B(2,2)$ 
are identical to those given in the appendix of \rcite{Huysmans2001} dividing by $1-\Delta$.
Elements $B(1,5)$, $B(2,5)$ and $B(5,5)$ are given by
\begin{equation}
 B(1,5) = i \rho_0 \frac{q R^2}{FF_M} \nabla \Psi \cdot \nabla \vartheta,
\end{equation}
\begin{equation}
 B(2,5) = \rho_0 \frac{qR^2}{fFF_M} |\nabla \Psi|^2,
\end{equation}
\begin{equation}
 B(5,5) = \rho_0 \frac{q R^2}{F_M} |B|^2.
\end{equation}

For \eqref{eq:sfl_momentum1}, 
the matrix elements $A^0(1,3)$, $A^0(1',3)$, $A^0(1,4)$, $A^0(1,4')$, $A^0(1',4)$ and $A^0(1',4')$
are same as those in the appendix of \rcite{Huysmans2001} and \cite{Chapman2006},
except that $dF/ds$ in \rcite{Huysmans2001} and \cite{Chapman2006} is now replaced by $\partial F/ \partial s$.
Other $A^0(i,j)$ elements coming from \eqref{eq:sfl_momentum1} are
\begin{eqnarray}
\fl A^0(1',6) =\frac{R^2}{fF_M}, \\
\fl A^0(1,6) = \frac{\partial}{\partial s} \left( \frac{R^2}{F_M} \right) \frac{1}{f} - A^0(1,7), \\
\fl A^0(1,7) = \frac{1}{f F_M B} \left( \frac{|\nabla \Psi|^2}{qB} \frac{dq}{ds} 
+ F \frac{\partial}{\partial s} \frac{BR^2}{F} 
+ f B F \frac{\partial}{\partial \vartheta} \frac{\nabla \Psi \cdot \nabla \vartheta}{F |B|^2} \right) \nonumber\\
+ i(m+nq)\frac{\nabla \Psi \cdot \nabla \vartheta}{F_M |B|^2}.
\end{eqnarray}

For \eqref{eq:sfl_momentum2}, the term $A^0(2,4)$ is same as \rcite{Huysmans2001}, 
but again changed its $dF/ds$ terms to $\partial F/ \partial s$.
Other elements are given by
\begin{eqnarray}
\fl A^0(2,3) = -\frac{1}{fqF}(m \bar{m} F^2 + n^2 q^2 |\nabla \Psi|^2) - \frac{(\bar{m} - m) m}{fq} F,\\
\fl A^0(2,4') = \frac{1}{fqF} (\bar{m}F^2 -nq |\nabla \Psi|^2) 
         + \frac{ \bar{m}-m} {fq} F,\\
\fl A^0(2,6) = \frac{m R^2}{fF_M} - A^0(2,7) ,\\
\fl A^0(2,7) =  \frac{i}{f F_M |B|^3}(|\nabla \Psi|^2 \partial_\vartheta B - F^2 \partial_\vartheta B + FB\partial_\vartheta F) \nonumber\\
 + (m+nq) \frac{|\nabla \Psi|^2}{f F_M |B|^2} .
 \end{eqnarray}
 
Also, $A^0(i,j)$ elements from right-hand side of Eq. (\ref{eq:sfl_momentum3}) are listed as following :
 \begin{eqnarray}
\fl A^0(5,3) = i (m+nq) \frac{F}{qR^2} \frac{\partial F}{\partial \vartheta} ,\\
\fl A^0(5,4) = \frac{m+nq}{q R^2} \frac{\partial |\nabla \Psi|^2}{\partial s} 
         + \frac{m+nq}{q^2R^2 F} |\nabla \Psi|^2 \left( F \frac{dq}{ds} - q \frac{\partial F}{\partial s} \right) \nonumber\\
         + (m+nq) \frac{fF}{qR^2} \frac{\partial}{\partial \vartheta} \frac{\nabla \Psi \cdot \nabla \vartheta}{F} 
         + (m+nq)  \frac{F}{q R^2} \frac{\partial F}{\partial s} \nonumber\\
         -i \frac{F}{q^2 R^2} \frac{\partial F}{\partial \vartheta} \frac{dq}{ds}  , \\
\fl A^0(5,6) = -i \frac{1}{(1-\Delta)B} \frac{\partial B}{\partial \vartheta} , \\
\fl A^0(5,7) = \frac{m+nq}{1-\Delta} - A^0(6,5).
\end{eqnarray}

The anisotropy related terms are given by
\begin{eqnarray}
\fl A^A(1,3) = \frac{1}{F_M} \left( \Delta  \beta_1 
+ R^2 \partial_s \Delta  \right) \left(m h_1 + n h_2 \right) \nonumber\\
+ 2n \frac{\nabla \Psi \cdot \nabla \vartheta}{F_M} \partial_\vartheta \Delta , \\
\fl A^A(1,4) =  \frac{\partial_\vartheta \Delta}{F_M} \left[ if(m+nq) \left( \frac{ F^2}{q^3 R^2 |\nabla \Psi|^2} 
+ \frac{\nabla \Psi \cdot \nabla \vartheta}{q |\nabla \Psi|^2} \right) 
- \frac{2 \nabla \Psi \cdot \nabla \vartheta}{q^2} \frac{dq}{ds} \right] \nonumber\\
+ \frac{\Delta}{F_M} [h_3 + ih_4 (m+nq) ] \beta_1 + \frac{\partial_s \Delta}{F_M} R^2 h_3, \\
\fl A^A(1,4') = \frac{\Delta}{F_M} h_5 \beta_1 
+ \frac{R^2}{F_M}h_5 \partial_s \Delta
+ 2 \frac{\nabla \Psi \cdot \nabla \vartheta}{q F_M} \partial_\vartheta \Delta,\\
\fl A^A(1',3) = \frac{\Delta R^2}{F_M} (m h_1 + n h_2), \\
\fl A^A(1',4) = \frac{\Delta R^2}{F_M} [h_3 + i(m+nq) h_4], \\
\fl A^A(1',4') = \frac{\Delta R^2}{F_M} h_5,\\
\fl A^A(2,3) = \frac{\Delta}{F_M} (m h_1 + n h_2) \beta_2
+ i \frac{R^2\partial_\vartheta \Delta}{F_M} (m h_1 - n h_2),\\
\fl A^A(2,4) = \frac{\Delta}{F_M} [h_3 + i(m+nq) h_4] \beta_2
- i \frac{R^2 \partial_\vartheta}{F_M} h_3 \nonumber\\
- (m+nq)\frac{|\nabla \Psi|^2 \partial_s \Delta}{fq F_M}, \\
\fl A^A(2,4') = \frac{\Delta}{F_M} \beta_2 h_5
+ i \frac{F^2 - |\nabla \Psi|^2}{fq F_M} \partial_\vartheta \Delta, \\
\fl A^A(5,3) = \frac{f \Delta}{1-\Delta} (m h_1 + n h_2) \beta_3
-i n \frac{|B|^2 \partial_\vartheta \Delta}{1-\Delta},\\
\fl A^A(5,4) = \frac{f \Delta}{1-\Delta} [h_3 + i (m+ nq) h_4] \beta_3
+ i \frac{ |B|^2 \partial_\vartheta \Delta}{q^2 (1-\Delta)} \frac{dq}{ds} \nonumber\\
- (m+nq) \frac{ |B|^2 \partial_s \Delta}{q (1-\Delta)}, \\
\fl A^A(5,4') = \frac{f \Delta}{1-\Delta} \beta_3 h_5
- i \frac{ |B|^2 \partial_\vartheta \Delta}{q (1-\Delta)},
\end{eqnarray}
in which 
\begin{eqnarray}
\fl \beta_1 =  - \frac{2R^2}{B} \frac{\partial B}{\partial s}
 - \frac{\partial R^2}{\partial s} + 2 \frac{R^2}{F} \frac{\partial F}{\partial s}
 - \frac{R^2}{F_M} \frac{dF_M}{ds} \nonumber\\
- 2 i f (\bar{m} + nq) \frac{ \nabla \Psi \cdot \nabla \vartheta}{|B|^2} 
 + 2 f \frac{\nabla \Psi \cdot \nabla \vartheta}{F|B|^2}  \frac{\partial F}{\partial \vartheta} \nonumber\\
 - \frac{2|\nabla \Psi|^2}{q|B|^2} \frac{dq}{ds},\\
\fl \beta_2 = 2i\frac{F}{|B|^2} \frac{\partial F}{\partial \vartheta}
- 2i \frac{R^2}{B}\frac{\partial B}{\partial \vartheta} - i \frac{\partial R^2}{\partial \vartheta}
+ \bar{m} R^2 \nonumber\\
- 2 (\bar{m} + n q) \frac{|\nabla \Psi|^2}{|B|^2}, \\
\fl \beta_3 = - i \frac{\partial_\vartheta \Delta}{(1-\Delta)^2} 
- 2 i \frac{1}{B} \frac{\partial B}{\partial \vartheta}
- (\bar{m} + nq),
\end{eqnarray}
and 
\begin{eqnarray}
\fl h_1 = -\frac{F^2}{fqR^2}, \quad
    h_2 =  \frac{|\nabla \Psi|^2}{fR^2}, \quad
    h_3 = -\frac{|\nabla \Psi|^2}{f q^2 R^2} \frac{dq}{ds}, \nonumber\\
\fl h_4 =  \frac{\nabla \Psi \cdot \nabla \vartheta}{qR^2},\quad
    h_5 =  \frac{|B|^2}{fq}.
\end{eqnarray}

\subsection{The ideal Ohm's law}
For the ideal Ohm's law equations (\eqref{eq:sfl_ohm1} and \eref{eq:sfl_ohm2}),
we have
\begin{eqnarray}
 B(3,3) = A^0(3,2) = 1, \\
 B(4,4) = -A^0(4,1)= 1,
\end{eqnarray}

\subsection{The single/double-adiabatic fluid closure equations}
The matrix element $B(6,6)$ and $B(7,7)$ are identical to \rcite{Chapman2006} $B(7,7)$.
For the single/double-adiabatic model \eqref{eq:sfl_pressure1} and \eref{eq:sfl_pressure2}, 
the $A^0(i,j)$ elements are given by
\begin{eqnarray}
\fl A^0(6,1') = - \gperone \frac{R^2}{F}, \\
\fl A^0(6,1) =  -\gperone \frac{\partial}{\partial s} \left(\frac{R^2}{F} \right) 
 - i f(\bar{m}+nq)\gperone \frac{\nabla \Psi \cdot \nabla \vartheta}{F |B|^2}   \nonumber\\
            - f \frac{\nabla \Psi \cdot \nabla \vartheta}{F |B|^2} \frac{\partial}{\partial \vartheta} (f_\perp-\gperone) 
            - \frac{R^2}{F} \frac{\partial f_\perp}{\partial s} \nonumber\\
           + \gpertwo \left[-\frac{|\nabla \Psi|^2}{F|B|^2} \frac{dq}{ds} - \frac{q}{B}\frac{\partial}{\partial s} \frac{BR^2}{F}
           +\frac{fq}{B} \frac{\partial}{\partial \vartheta} \frac{\nabla \Psi \cdot \nabla \vartheta}{BF} \right],
\end{eqnarray}
\begin{eqnarray}
\fl A^0(6,2) =  \gperone \frac{F}{|B|^2} (n \frac{q|\nabla \Psi|^2}{F^2} - \bar{m}) 
    + i \frac{F}{|B|^2} \frac{\partial}{\partial \vartheta} (f_\perp-\gperone) \nonumber\\
    + i \gpertwo \frac{1}{B} \frac{\partial}{\partial \vartheta} \frac{F}{B},
\end{eqnarray}
\begin{eqnarray}
\fl A^0(6,5) =  - \gperone (\bar{m}+nq) - \gpertwo(m+nq) + i \frac{\partial}{\partial \vartheta} (f_\perp-\gperone) \nonumber\\
           +i \gpertwo \frac{1}{B} \frac{\partial B}{\partial \vartheta},
\end{eqnarray}
Replacing $f_\perp$ by $f_\parallel$, $\gperone$ by $\gparone$ and $\gpertwo$ by $\gpartwo$,
we will reach the matrix elements $A^0(7,1')$, $A^0(7,1)$, $A^0(7,2)$ and $A^0(7,5)$.

\subsection{The incompressible fluid closure }
The matrix element $B(6,6)$ and $B(7,7)$ are identical to \rcite{Chapman2006} $B(7,7)$.
The $A^0(i,j)$ elements originated from \eqref{eq:pressure_IC1} and \eref{eq:pressure_IC2}
are given by
\begin{eqnarray}
A^0(4,1) = - \frac{R^2}{F} \left( \frac{\partial \pper}{\partial s}  
 - \frac{\partial_\vartheta \pper}{\partial_\vartheta B} \frac{\partial B}{\partial s} \right), \\ 
A^0(7,1) = - \frac{R^2}{F} \left( \frac{\partial \ppar}{\partial s}  
 - \frac{\ppar-\pper}{B} \frac{\partial B}{\partial s} \right).
\end{eqnarray}

\section*{References}
\bibliography{anisotropy_sta}
\end{document}